\documentclass[%
reprint,
superscriptaddress,
amsmath,amssymb,
aps,
]{revtex4-2}

\usepackage{graphicx}
\usepackage[table]{xcolor}
\usepackage{dcolumn}
\usepackage{bm}

\newcommand{\samresponse}[1]{}

\graphicspath{{./../format_pnas_figures/}}
\graphicspath{{./finalfigs/}}

\begin{document}

\title{Diffusive Spreading Across Dynamic Mitochondrial Network Architectures}

\author{Keaton B. Holt}
\affiliation{Department of Physics, University of California, San Diego, San Diego, California 92093}
\author{Camryn Zurita}
\affiliation{Department of Molecular and Cell Biology, University of California, Berkeley, Berkeley, California 94720}
\author{Lizzy Teryoshin}
\affiliation{Department of Physics, University of California, San Diego, San Diego, California 92093}
\author{Samantha C. Lewis}
\affiliation{Department of Molecular and Cell Biology, University of California, Berkeley, Berkeley, California 94720}
\affiliation{Innovative Genomics Institute, Berkeley, California 94720}
\affiliation{Helen Wills Neuroscience Institute, Berkeley, California 94720}
\author{Elena F. Koslover}
\email{ekoslover@ucsd.edu}
\affiliation{Department of Physics, University of California, San Diego, San Diego, California 92093}

\date{\today}

\newcommand{\change}[1]{{#1}}

\begin{abstract}
	\change{
		In eukaryotic cells, mitochondria form networks that range from highly fused interconnected structures to fragmented populations of individual organelles that undergo transient interactions. These structures can be described as temporal networks of physical units, whose dynamic topology is determined by fusion, fission, and motion of the mitochondria through intracellular space. The heterogeneity of the mitochondrial population is governed by diffusive transport and inter-unit exchange of proteins, lipids, ions, and RNA within these networks. We present a unifying framework for 
		the dispersion of material within temporal networks of spatially embedded units that span across a broad connectivity range. Specifically, we consider filling of the networks with a locally produced but globally consumed material, demonstrating that the steady-state content is determined by the balance of timescales for spatial encounter between clusters, local fusion, fission, and diffusive transport within a cluster. 
		As the connectivity increases, filling behavior transitions from three-dimensional spread through a `social network' limited by cluster interactions to low-dimensional transport through a largely stationary `physical network' limited by material diffusivity. We extract parameters for mitochondrial networks in three human cell lines, demonstrating that different cells can access both the social and the physical network regimes. These results provide a quantitative basis for predicting the homogenization of biomolecules through a mitochondrial population. Our framework unifies a variety of temporal network structures into an overarching theory for transport through populations of interacting and interconnected units.
	}
\end{abstract}

\maketitle
%

Diffusive transport through networks has been studied in a variety of contexts, including disease spread in epidemiology~\cite{tatem2006global,newman2002spread,peruani2008dynamics,rodriguez2022epidemic},  innovations in social networks~\cite{montanari2010spread},
communication among insects~\cite{gernat2018automated,blonder2011time}, 
diffusion of signals and nutrients in the brain extracellular space~\cite{kinney2013extracellular,sykova2008diffusion}, and oil recovery in porous rock~\cite{king1999predicting}. These systems can be separated into two broad classes: one where the topology of network connections is stationary and limited to nearest neighbors, and another where network nodes are mobile and promiscuous, interacting with many different partners over time~\cite{rodriguez2022epidemic}.

Physical networks form an important category of \change{topologically} stationary, spatially-constrained network structures~\cite{barthelemy2011spatial}. In such networks, edges represent objects subject to physical limitations (space-filling, steric repulsion, etc.)~\cite{dehmamy2018structural,posfai2024impact} connecting 
degree-1 tips and degree-3 junctions,
 with higher degree nodes exceedingly rare~\cite{blagojevic2024three,sukhorukov2012emergence}. Examples  include fungal mycelia~\cite{islam2017morphology}, neuronal synaptic networks~\cite{scheffer2020connectome}, and interconnected pores in rocks~\cite{ju20143d}. Transport behaviors on stationary networks can be described by the graph Laplacian, whose eigenvalues govern the spreading timescales~\cite{chung1997spectral,masuda2017random,pete2024physical}.
In some percolation problems, physical networks become dynamic as edges are allowed to flicker between an active and inactive state~\cite{hoitzing2015function,hermon2020comparison,chuphal2024mitochondrial}. \change{However, the set of possible connections for each node remains tightly limited to its physically proximal neighbors.}

A distinct set of approaches considers signal transmission in `social' networks of transiently-interacting units.
In such networks, 
\change{spreading is governed}
 by distributions of contact duration and inter-contact times~\cite{gernat2018automated,holme2012temporal}, which may be constrained by spatial embedding~\cite{newman2002spread,tatem2006global}. \change{Models that use locally-interacting mobile agents moving through two-dimensional space have been shown to accurately represent topological properties for human friendship and sexual interaction networks~\cite{gonzalez2006system}.
Both the contact rate and the probability of transmission during each contact may depend on the particle mobility, resulting in multiple distinct scaling regimes as the mobility is increased~\cite{peruani2008dynamics}.}
Notably, many previously studied social network systems focus on the spreading of a non-diluting signal (infection, information, etc.)~\cite{gernat2018automated,newman2002spread,montanari2010spread,peruani2008dynamics,rodriguez2022epidemic}, in contrast to dispersion of mass-conserving physical material~\cite{herrera2013fractal,chuphal2024mitochondrial,scott2023endoplasmic}.

Despite the distinct modeling approaches employed, physical and social  networks lie on a continuum of temporal network structures~\cite{holme2012temporal} with varying timescales of topological rearrangement. \change{Understanding how the physical behavior of moving agents translates into temporal network properties and the transmission of signals upon such networks remains an open question in the field. Recent results have highlighted the importance of relative timescales for network rearrangement versus local transmission, with infection spread transitioning from a static locally-connected regime to a homogeneously mixed mean-field regime as the particle mobility increases~\cite{rodriguez2022epidemic}.}	

The intracellular environment exemplifies the full spectrum of temporal network regimes. For example, the endoplasmic reticulum forms a well-connected physical lattice of membrane-bound tubules that  enables diffusive transport of proteins~\cite{scott2023endoplasmic} and ions~\cite{crapart2024luminal} across the entire cell. At the opposite extreme, the population of endocytic vesicles constitutes a social network of discrete transiently-interacting compartments~\cite{york2023deterministic,cason2022spatiotemporal}.

Mitochondria form another intracellular network with striking structural variability. Mitochondrial network architectures vary across cell types~\cite{wang2023mitotnt,viana2020mitochondrial,aon2004percolation,chustecki2021network}, and transition between fragmented and hyperfused states in response to disease~\cite{chen2009mitochondrial,burte2015disturbed,wang2020mitochondria,park2018mitochondrial,filosto2011role,izzo2017metformin}, metabolic conditions~\cite{liesa2013mitochondrial,rambold2011tubular,kichuk2024using}, calcium signaling and apoptosis~\cite{suen2008mitochondrial,hom2007thapsigargin}, and cell cycle progression~\cite{mishra2014mitochondrial,mitra2009hyperfused}. \change{Recent studies have highlighted the importance of mitochondrial dynamics in allowing the cell to tune between conflicting functional demands~\cite{chustecki2024collective,hoitzing2015function}. On the one hand, maintaining a broad distribution of mitochondria throughout the cell helps limit ATP gradients~\cite{schuler2017miro1,kumar2025estimating}, facilitates the formation of contact sites with many other organelles~\cite{prinz2020functional}, and allows mitochondria to serve as local calcium reservoirs and signal transduction hubs~\cite{giorgi2018machineries,picard2022mitochondrial}. On the other hand, fusion into more compact interconnected structures may contribute to reduction of mitochondrial heterogeneity~\cite{ngo2021mitochondrial,ryu2024cellular}, genetic complementation~\cite{chen2010mitochondrial,kowald2011evolution,giannakis2022exchange}, quality control of the mitochondrial population~\cite{kleele2021distinct,patel2013optimal,kim2007selective,twig2008fission}, dilution of harmful reactive oxygen species (ROS)~\cite{yu2006increased}, and power cabling for enhanced ATP production~\cite{hoitzing2015function,skulachev2001mitochondrial}. Rapid transport and transient fusion between small mitochondrial units, as observed in plant cells~\cite{chustecki2021network,giannakis2022exchange, chustecki2024collective}, enable broad dispersion of mitochondria throughout the cellular space while also allowing some mixing of mitochondrial contents within the population. The homogenization of mitochondrial material relies both on diffusive spreading within connected networks~\cite{brown2020impact,chuphal2024mitochondrial} and on dynamic encounters and exchange events between disjoint clusters~\cite{chustecki2024collective}. Establishing the functional consequences of different mitochondrial morphologies thus requires an overarching picture of the relative contributions from these distinct dynamic processes.
}

\change{To bridge the gap between mitochondrial structure and material transport, we present a quantitative framework for diffusion on spatially embedded dynamic networks. Our central finding is a connectivity-driven transition in transport: from low-dimensional tunneling through well-connected networks to  three-dimensional dispersion across a population of transiently interacting units. 
We present a unifying expression that approximates the rate of network filling from a source across multiple distinct regimes, tuned by the interplay of competing timescales: the cluster filling rate, encounter rate, fusion rate, and the decay rate of the spreading material. The model is applied to  mitochondrial networks in three human cell types, demonstrating the structural diversity and different regimes of relevance for these networks.}

\begin{figure}
	\includegraphics[width=8.3cm]{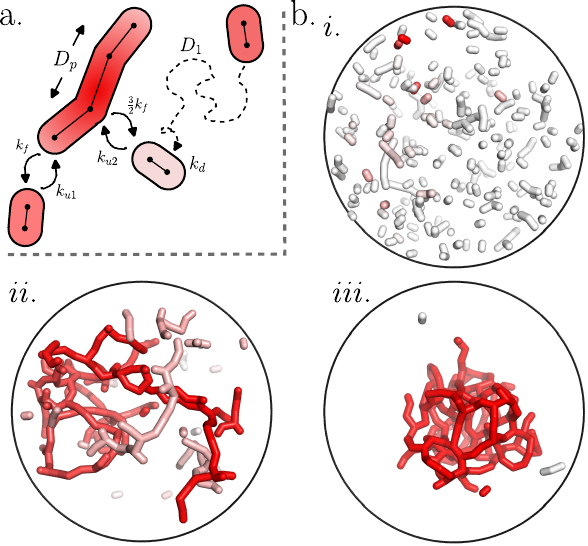}
	\caption{Dynamic network simulation framework. (a) \change{Schematic of network formation} by interacting units which diffuse through 3D space with diffusivity $D_1$. Nearby units can undergo tip-tip and tip-side fusion with rates $k_{u1},k_{u2}$ while connected nodes undergo fission at rate $k_f, \frac{3}{2}k_f$, respectively. Material spreads along connected units with diffusivity $D_p$ and decays over time with rate $k_d$. Red indicates the concentration of material in a given unit. (b) \change{Snapshots of network filling from a single source unit, at steady state.}
		 (i) Fragmented network, ($k_{u1}/k_f = 30$) (ii) Network near percolation transition, with kinetic parameters appropriate to mammalian mitochondrial networks~\cite{holt2024spatiotemporal} ($k_{u1}/k_f = 1000 $). (iii) Hyperfused network ($k_{u1}/k_f = 3000$). The ratio of tip-tip and tip-side fusion is set to $k_{u1} = 3k_{u2}$, the dimensionless particle diffusivity is $D_p = 4800$, \change{and the dimensionless material decay rate is $k_d=0.32$ throughout}.}
	\label{fig:intro_diagram}
\end{figure}

\section*{\label{sec:problem_statement} Model description}
We seek to quantify the rate of material dispersion through dynamic spatial networks ranging from the social to the physical regime. These limiting architectures, as well as a continuum of intermediate network connectivities, are encompassed by a simulation framework~\cite{holt2024spatiotemporal}, illustrated in Fig.~\ref{fig:intro_diagram}, consisting of $N_0$ interacting spherocylindrical units of length $\ell_0$. 
\change{Each node (unit endpoint) is subject to random, uncorrelated Brownian forces, resulting in an effective translational diffusivity $D_1$ for individual units. The nodes are also subject to elastic bending, stretching, and steric repulsion forces~\cite{holt2024spatiotemporal}, analogous to standard polymer mechanics models.}
Upon close encounters, pairs of units may fuse into larger clusters with rate constants $k_{u1}$ (for tip-tip fusions) or $k_{u2}$ (for tip-side fusions). 
Fission breaks connections at the nodes, with fission rate proportional to the number of attached edges ($k_f$ at degree-2 nodes and $1.5 k_f$ at degree-3 nodes), in keeping with prior models~\cite{sukhorukov2012emergence}. \change{The unit length $\ell_0$ thus represents the minimal cluster size that can be formed by fission and the minimal separation between junctions.}
 At steady state, the system exhibits a characteristic mean cluster size $\langle n \rangle$ set by the balance of fusion and fission, and an associated \change{translational} cluster diffusivity $D_n$. 
 
 \change{We note that this structural model differs from classic diffusion-limited aggregation or cluster-cluster aggregation models~\cite{krapivsky2010kinetic,meakin1984topological} in that the degree at each node is limited, there are mechanical moduli that constrain bending at linear nodes and junctions, and the fusion rate is dependent on angle and node degree. The spherocylindrical units enable explicit incorporation of these effects to represent the physical features of mitochondrial networks, including separately tuning the length of snake-like segments between junctions and the cluster size of the networks.
 Details of the structural model and the selection of appropriate parameters are described in Ref.~\cite{holt2024spatiotemporal}, and summarized in the SI Appendix.} Here we focus on the spreading of material throughout the resulting dynamic networks. 

\change{The network structures are allowed to run to steady state prior to the start of the material spreading simulation.}
We then select at random a single unit, which serves as a particle source with fixed constant concentration $h_0=1/\ell_0$. Such a fixed source could represent, for instance, an import site for nucleus-encoded proteins~\cite{khan2024mitochondrial}, a site of regulated translation at a mitochondrial RNA granule or nucleoid~\cite{landoni2024mitochondrial}, or a location of calcium entry at a contact with the endoplasmic reticulum~\cite{csordas2010imaging}.
The concentration field is propagated between connected units on the network with effective diffusivity $D_p$, using a finite volume approach~\cite{leveque2002finite} discretized at the level of individual units. Concentrations decay with a constant rate $k_d$ (representing removal from the network).
The decay rate serves to set a relevant timescale, with the resulting calculations closely related to the question of how much material spreads from the source over a certain time. 
We quantify dispersion by computing the total amount of material ($S$) in the network at steady-state, excluding the source unit. Example steady-state snapshots are shown in Fig.~\ref{fig:intro_diagram}b and Supplemental Video 1.

Given the complex internal structure of mitochondria and the variety of functionally relevant biomolecules within them, particle diffusivities can vary broadly. For example, \change{proteins in the inner mitochondrial membrane can be embedded in the extensive folds of cristae, resulting in much slower diffusivity than the corresponding outer membrane proteins~\cite{sukhorukov2009anomalous,jakubke2021cristae}.} The cristae also serve as diffusive barriers to hinder transport of solutes in the matrix~\cite{dieteren2011solute} and intermembrane space~\cite{rampelt2017role}.
Our simplified modeling approach coarse-grains these complications into a single effective diffusivity $D_p$ representing the motion of a particle along the mitochondrial tubule axis.
 Reported effective diffusivities vary over orders of magnitude, from
$\sim0.004\mu\text{m}^2/\text{s}$ for ATP synthase components ~\cite{sukhorukov2010determination}, to $\sim20\mu\text{m}^2/\text{s}$ for mitochondrial matrix proteins~\cite{koopman2008inherited}, with diffusivity of ions presumed even higher~\cite{gerencser2005mitochondrial}. We consider a range of diffusivities $D_p= 0.4-40\mu\text{m}^2/\text{s}$ in our simulations. 
\change{The averaged diffusivity of individual mitochondria, which includes occasional active transport along microtubules,}
is estimated to be much slower, at $D_1 \approx 0.25\mu\text{m}^2/\text{min}$~\cite{wang2023mitotnt,corci2023extending}.

The results below are reported in dimensionless units, relative to the length scale $2\ell_0$ and the timescale $1/k_f$. A reasonable estimate sets the mitochondrial unit length to $\ell_0 = 0.5\mu\text{m}$ and the fission timescale to $1/k_f \approx 2 \text{ min}$~\cite{holt2024spatiotemporal}, consistent with experimental measurements of \change{fragmented mitochondrial volume~\cite{rafelski2012mitochondrial,kaasik2007regulation,jakobs2014super} and} overall fission rate~\cite{twig2010biophysical,wang2023mitotnt}. The corresponding dimensionless diffusivities are $D_p = 48 - 4800$ for the material, and $D_1 = 0.5$ for individual mitochondria.

\begin{figure*}
	\includegraphics[width=17.3cm]{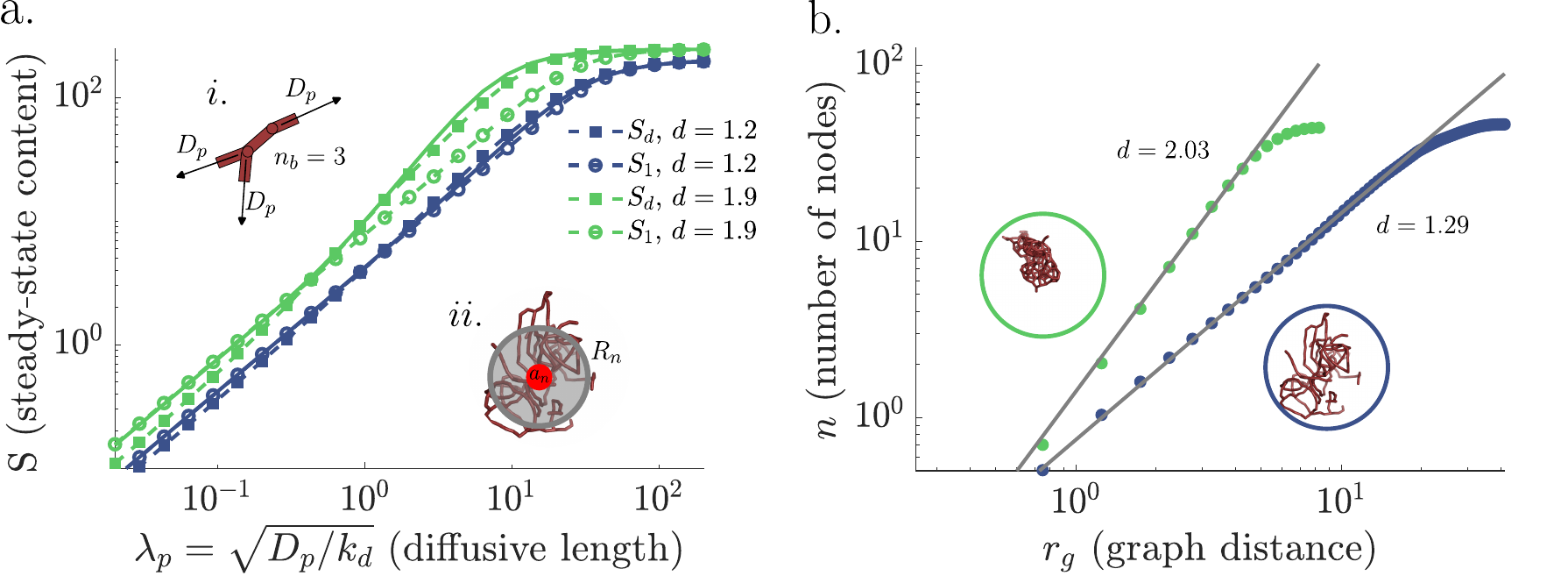}
	\caption{
		\change{Filling of static networks depends on network dimensionality.} Two different sets of networks are considered:  near-linear structures (blue) with $k_{u2} = 0.01k_{u1}$ and highly-branched structures (green) with $k_{u2} = 10k_{u1}$. The simulations are run to steady state, and the network structures are then frozen. 
		\change{(a) Steady-state material content supplied by a }
		source unit on the static networks is plotted as a function of 
		the diffusive lengthscale, $\lambda_p=\sqrt{D_p/k_d}$. Solid curves show exact solution (details in SI Appendix), averaged over replicate snapshots. 
		Dashed-circle curves show approximation with the linear motif (inset i), applicable at short $\lambda_p$. Dashed-square curves show continuum solution on a fractal domain (inset ii), applicable at long $\lambda_p$.
		(b) Calculation of the graph dimension for a single instance of each network type (shown in inset). The number of nodes within a given graph distance is plotted against the graph distance on log-log axes, with the slope giving the intrinsic dimension $d$.  Averaging over 21 snapshots from 3 independent simulations yields effective dimensionalities $d = 1.2, d=1.9$, respectively. 	 	
	}
	\label{fig:ddimmatchescontinuum}
\end{figure*}

\section*{\label{sec:results} Results}

Spreading on the network is governed by the interplay of several key timescales: the decay time $\tau_\text{d} = 1/k_d$, the cluster filling time $\tau_\text{c}$ for particles to diffusively explore a typical-sized connected component (cluster), the waiting time between interactions $\tau_\text{int}$ (the time to encounter and fuse with a new cluster), and the fission time $\tau_f = 1/k_f$ which sets the typical duration of transient interactions between clusters.
The material exhibits different dynamics depending on the comparative values of these four timescales. We begin by considering limiting regimes where different timescales dominate.

\subsection*{Highly connected regime: spreading through stationary networks}
In the limit where $\tau_d$ is the shortest timescale, material spreads primarily within a single connected component. This regime is relevant for highly fused networks with large cluster sizes. Such networks tend to be relatively static in their topology, with any fissions that occur rapidly followed by re-fusion with the same neighboring unit~\cite{holt2024spatiotemporal}. Recent evidence of colocalizing mitochondrial fusion and fission machinery at ER-mitochondria contact sites also points towards increased likelihood of re-fusion following a fission event~\cite{abrisch2020fission}.

For static networks, the steady-state distribution of material can be found by solving the diffusion equation on each one-dimensional edge, while matching boundary conditions at each node (details in SI Appendix). This method yields an exact solution for each individual network (Fig.~\ref{fig:ddimmatchescontinuum}a). \change{However, simplified approximations of the network structure provide further insight on the relevant scaling regimes.}

Depending on the relative rates of tip-tip versus tip-side fusion, network architectures can range from near-linear snake-like structures to highly branched compact morphologies, with the latter allowing for more rapid spreading of material through the network (Fig.~\ref{fig:ddimmatchescontinuum}). We first consider the snake-like limit, relevant when the
 diffusive length scale $\lambda_p = \sqrt{D_p/k_d}$ is smaller or comparable to the length of linear segments between junctions. In the case of a completely linear network, the steady-state material content $S_1$ is given by the solution of the one-dimensional (1D) diffusion equation, $S_{1}=\frac{n_b\lambda_p}{\ell_0}\tanh \left( \frac{\ell_0\left<n\right>/n_b}{\lambda_p}\right)$
with $n_b = 2$. \change{Moving beyond the linear limit, additional branches adjacent to the source enable more material to enter the network.} We thus define $n_b$ as the average number of edges directly connected to the source unit, and approximate the entire network cluster as a set of $n_b$ linear spokes, each of length $\ell_0\left<n\right>/n_b$, connected to a central source (Fig.~\ref{fig:ddimmatchescontinuum}
\change{$i$}). $S_1$
then gives an estimate of the total material spreading over a short diffusive length-scale $\lambda_p$ (Fig.~\ref{fig:ddimmatchescontinuum}\change{a}).

\change{For the regime where material spreads over a more extensive network structure ($\lambda_p$ much greater than segment length), we turn to a fractal continuum approach.  
Complex and porous media are often described as self-similar fractal structures characterized by polynomial scaling laws that relate different measures of distance and system size~\cite{stauffer2018introduction,stanley1984application,yu2004fractal}. 
 For our simulated network structures, we count the number of nodes within a given graph distance from a starting node, averaged over all possible starting nodes~\cite{shanker2007defining}. 
 The power-law scaling exponent of this curve (Fig.~\ref{fig:ddimmatchescontinuum}
\change{b}) \change{defines  the intrinsic fractal dimension $d$ of the network}. Sometimes called $d_\ell$, the `chemical' or `graph' dimension~\cite{havlin1985chemical,porto1997structural},  this value describes the topology of the network and is distinct from the extrinsic dimension describing how the network is laid out in space~\cite{stauffer2018introduction,ben2000diffusion}.}

\change{
Although the simulated networks are embedded in  three-dimensional space, their graph dimension is substantially lower ($1\le d \le 2$).
 This feature has been previously observed in a variety of physical networks, from plant roots to ant tunnels, composed of tube-like objects connected at junctions~\cite{blagojevic2024three}. Three-dimensional percolation networks, often used to describe transport through porous media~\cite{sahimi1993flow}, have a graph dimension of $d \approx 1.8$~\cite{stauffer2018introduction,ben2000diffusion}. Structures arising from irreversible cluster-cluster aggregation have $d \approx 1.4$~\cite{meakin1984topological}. These low dimensionalities imply that the tortuosity of paths~\cite{ghanbarian2013tortuosity,fu2021tortuosity} (ratio of path length to Euclidean distance) increases with system size, leading to effectively subdiffusive motion through space for particles embedded within the network~\cite{ben2000diffusion}. 
For a network of $n$ units, with intrinsic graph dimension $d$, the cluster filling timescale can be estimated as: $\tau_c= (\left<n\right>^{1/d} \ell_0)^2/D_p$.
}

To compute the steady-state material content, we approximate the network as a hyperspherical domain of dimension $d$, 
 with a sphere of radius $a_n$ (enclosing the source edge) maintained at fixed concentration, and a reflecting sphere of radius $R_n$ representing the outer boundary of the cluster \change{(Fig.~\ref{fig:ddimmatchescontinuum}$ii$)}. We set $R_n$ such that the average distance between two points in the $d$-dimensional domain is equal to the average graph distance between network nodes (details in SI Appendix). 
Assuming that the $\left<n\right>$ network units are uniformly distributed within the continuum sphere, the radius $a_n$ is set proportionately to allow for 1 source unit within the inner sphere: $1/a_n^d = (\langle n \rangle-1)/(R_n^d-a_n^d)$.

\change{Assuming the system is spherically symmetric, we can write down the steady-state diffusion equation in $d$-dimensional space~\cite{redner2001guide} in terms of the radial coordinate $r$, as 
$\left(\frac{1}{d}D_p\right)\frac{1}{r^{d-1}}\frac{\partial}{\partial r}\left(r^{d-1}\frac{\partial c(r)}{\partial r} \right)-k_dc(r)=0$. 
The effective particle diffusivity is scaled by the dimension $d$ because,
when diffusing along an edge, the particle only moves in one dimension at each instance in time rather than simultaneously in all $d$ dimensions.}
The $d$-space diffusion equation has solutions in terms of modified Bessel functions $I_\nu(x), K_\nu(x)$~\cite{redner2001guide},
where $\nu = 1-d/2$ and $x = r\sqrt{d}/\lambda_p$.
Taking a fixed concentration boundary at $x_1=a_n\sqrt{d}/\lambda_p$ and a reflecting boundary at $x_2=R_n\sqrt{d}/\lambda_p$ we find the 
\change{steady-state material content}
(details in SI Appendix):
\begin{equation}
\begin{gathered}
S_{d}=\frac{d}{x_1}\frac{\left[ -I_{\nu-1}(x_1)K_{\nu-1}(x_2)+I_{\nu-1}(x_2)K_{\nu-1}(x_1) \right]}{\left[ I_\nu(x_1)K_{\nu-1}(x_2) +I_{\nu-1}(x_2)K_\nu(x_1)\right]} \\
\end{gathered}
\label{eq:ddimsoln}
\end{equation}

Fig.~\ref{fig:ddimmatchescontinuum}\change{a} shows the correspondence between diffusion through a fractal continuum and the exact solution for networks with different connectivities. 
The linear motif solution $S_1$  is a good estimate for small diffusive lengths, while the continuum solution $S_d$ is a better approximation for large $\lambda_p$. \change{This transition is analogous to a `representative elementary volume' used to describe the minimal size of porous media above which a self-similar fractal approximation can apply~\cite{hill2013free,fu2021tortuosity}.}
When the diffusive length-scale is large enough ($\lambda_p > R_n$), the entire connected component saturates. Overall, material delivery is boosted by increased network branching~\cite{chuphal2024mitochondrial}, which increases both the network dimension and the network density (lower $R_n$ for the same number of network units).

\subsection*{Fragmented network regime: spreading through transient interactions}
We next consider the regime of highly fragmented `social' networks, where clusters are sufficiently small (and particle diffusivity sufficiently fast) that concentrations are fully equilibrated within each cluster: $\tau_c \ll \left\{\tau_\text{d}, \tau_\text{int} \right\}$. In this limit, we can approximate the  network as a system of $N=N_0/\left<n\right>$ identical spherical units with effective radius $a$, each representing a cluster of uniform concentration. These effective units are capable of undergoing transient fusions whenever they are within contact radius $b$, with local rate constant 
$k_u$. Each fusion equilibrates the particle concentration in the two units involved, and is instantaneously followed by fission so that no larger structures are formed. 

\begin{figure}
	\includegraphics[width=8.3cm]{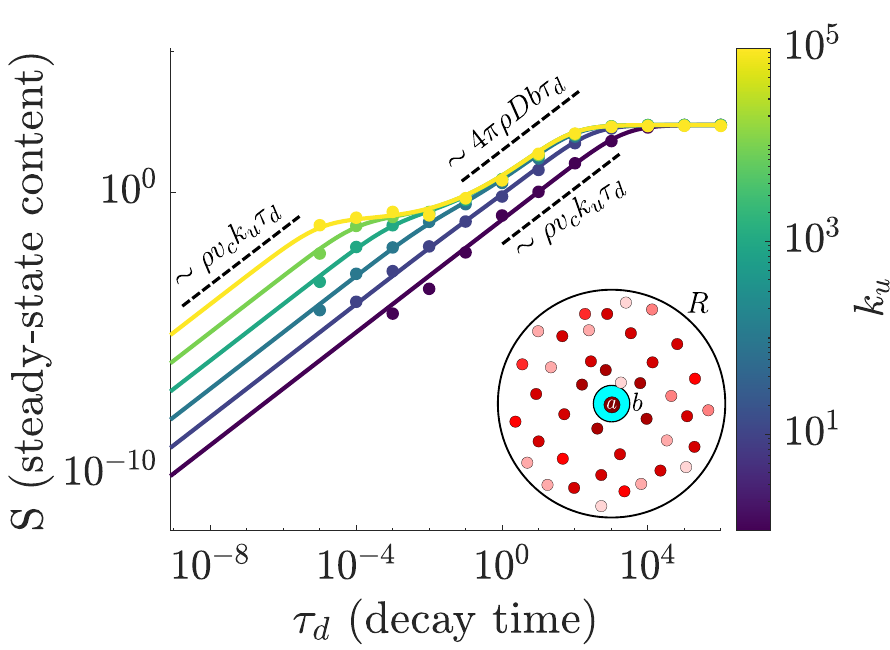}
	\caption{
	\change{Steady-state material content supplied by a}
	source unit in a system of fragmented clusters. 
		 Solid lines show mean-field solutions (Eq.~\ref{eq:meanfieldsoln}) with unit cluster size $\langle n \rangle=1$, for different fusion rates $k_{u}$. Colored dots show explicit simulation results for a simplified system of interacting spheres with uniform size (inset).
		Dashed black lines show the limiting behavior for fusion-limited and encounter-limited regimes. }
	\label{fig:meanfieldsoln}
\end{figure}

Results from an explicit simulation of this simplified model are shown in Fig.~\ref{fig:meanfieldsoln}. A mean-field analytic approximation can be found by fixing the source unit in the center of a domain, assigning a relative diffusivity of $D=2D_n$ to the remaining units and solving for the spatial field $h(r,t)$, which defines the \change{material per network unit} located at distance $r$ from the origin. The average of this field within the narrow contact zone is defined as $h_c(t)$. The time evolution of the concentration fields can be expressed as follows (details in SI Appendix):
\begin{subequations}
	\begin{align}
		\frac{\partial h(r,t)}{\partial t} & = D\nabla ^2 h(r,t)-k_dh(r,t),  \;\; \text{for}\ b<r<R \\
	\frac{dh_c(t)}{dt} & = k_u\left[h_0\ell_0-h_c(t)\right] - \frac{I}{\rho v_c} -k_dh_c(t),
	\end{align}
	\label{eq:meanfieldodes}%
\end{subequations}
where $k_u\left[h_0\ell_0-h_c(t)\right]$ represents the injection of new material into the system via fusion with the source, $I= -4\pi b^2\rho D\left.\frac{\partial h(r,t)}{\partial r}\right|_b$ is the current of material leaving the contact zone,  $\rho= (N_0-\langle n\rangle)/V$ is the density of network units, \change{$V = \frac{4}{3} \pi (R^3-a^3)$} is the domain volume, and \change{$v_c=\frac{4}{3} \pi (b^3-a^3)$} is the contact zone volume. Interactions between non-source units do not alter the mean-field concentrations, and the overall material in the system can only increase upon encounters with the source.

We compute the steady-state solution of Eq.~\ref{eq:meanfieldodes} and integrate $h(r)$ over space to find the total amount of material ($S$) in the system, excluding the source unit (details in SI Appendix):
\begin{subequations}
	\begin{align}
		S 
	= & h_0\ell_0(\langle n \rangle - 1) + h_c \rho (v_c+z),
		\\
		h_c = & \frac{h_0\ell_0}{1+k_d/k_u+k_dz/(k_uv_c)}, \\
		\frac{z}{4\pi b\lambda} = & \frac{(\lambda-b)(R+\lambda)+(\lambda+b)(R-\lambda)e^{2(R-b)/\lambda}}{(R+\lambda) + (R-\lambda)e^{2(R-b)/\lambda}}, 
	\end{align}
	\label{eq:meanfieldsoln}%
\end{subequations}
where $\lambda = \sqrt{D/k_d}$ is the diffusive lengthscale for spreading through the population of clusters.
Here, the first term in $S$ represents material in the  cluster containing the source unit, the term $h_c \rho v_c$ is the total material within the contact layer (defined by a balance of fusion, decay, and diffusive escape), and $z$ describes the additional volume over which the material has spread beyond the contact zone.

The total material in the system exhibits distinct scaling behaviors (Fig.~\ref{fig:meanfieldsoln}) depending on the relative timescales for decay, encounter ($\tau_\text{enc} = (4\pi D b \rho)^{-1}$), and fusion ($\tau_u = (k_u \rho v_c)^{-1}$). In keeping with standard results for diffusion to a partially reactive target~\cite{logan1967effects}, the overall interaction time is given by the sum of waiting times for the two-step process: $\tau_\text{int} = \tau_\text{enc} + \tau_u$. 
In the limit of fast decay ($\tau_d \ll \tau_\text{int}$), the steady-state material in a social network of unit-size clusters approaches
$S \rightarrow \rho v_ch_c = \rho v_ck_u/k_d$.
In this limit, the \change{steady-state content} is set by the number of units in the contact zone ($\rho v_c$) and how often they fuse with the source during the decay time ($k_u/k_d$), with no spatial spread.

Another limit arises when diffusion of clusters is fast relative to both decay and fusion
 ($\tau_\text{enc} \ll \tau_d, \tau_u$).
 This yields the same scaling for the \change{steady-state} content $S\rightarrow \rho v_ck_u/k_d$, with \change{$h_c \rightarrow k_uv_c/(4\pi Db)$}.
The amount of material in the contact layer ($h_c$) is set by a balance between injection through fusion (at rate $k_u \rho v_c$) and escape through the diffusive arrival of fresh units that dilute the local concentration ($4\pi Db\rho$). The additional volume $z\rightarrow  4\pi Db / k_d$ over which the material spreads depends on the balance between diffusive encounters and decay. Overall, the total amount of material in the network is fusion-limited as the rapid encounters quickly homogenize the individual units within the diffusive range of the source. \change{We note that this limit is analogous to the homogeneous-mixing limit described for epidemic spread among rapidly diffusing mobile agents~\cite{rodriguez2022epidemic}.}

In a third limit, fusion of clusters is fast and decay is slow relative to the timescale of diffusive encounter. ($\tau_u \ll \tau_\text{enc} \ll \tau_d$). For this regime, the total material content is given by
$S\rightarrow \rho zh_c \rightarrow 4\pi Db\rho/k_d$.
This is an encounter-limited regime, where clusters arriving at the contact region fuse with the source nearly instantaneously ($h_c\approx h_0\ell_0=1$), and spreading is determined by the balance of arrival and decay rates.

Fig.~\ref{fig:meanfieldsoln} shows the full solution of the mean-field model (Eq.~\ref{eq:meanfieldsoln}) for spreading on a fragmented network, with variable fusion rate $k_u$ and a unit cluster size $\left<n\right>=1$.
 The analytic calculations accurately reproduce simulations with diffusive spheres that exchange material via transient fusion (details in SI Appendix).

\change{For simplicity, our model assumes effectively diffusive motion of mitochondrial units. However, individual mitochondria in both plant and animal cells can also engage in directed motor-driven runs~\cite{chustecki2021network,giannakis2022exchange,winter2025asymmetric}. Such motion increases the effective diffusivity of the mitochondria over timescales encompassing multiple runs, and concomitantly increases the encounter rate between particles~\cite{rodriguez2022epidemic, teryoshin2025encounter}. The effect of processive runs on network filling in the fragmented regime is shown in Supplemental Fig.~S1 in the SI Appendix. As reported previously, long runs allow for more rapid mixing of network contents~\cite{giannakis2022exchange}, with diminishing effect once the run-length becomes comparable to the domain size~\cite{teryoshin2025encounter,rodriguez2022epidemic}.}

\begin{figure*}
	\includegraphics[width=17.3cm]{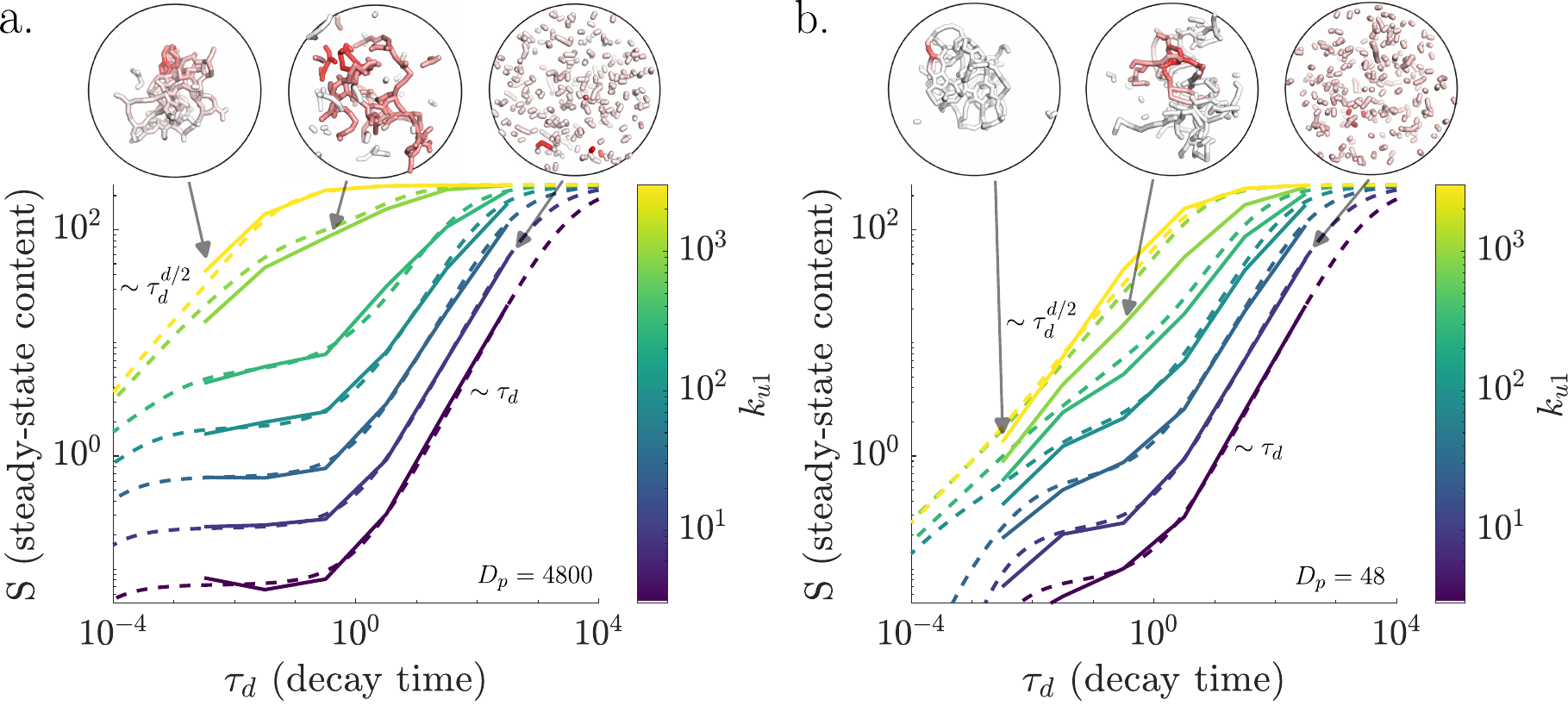}
	\caption{Spreading on simulated networks exhibits a transition between the static network and social network regime.		
	\change{The steady-state material content} is plotted as a function of the decay time for different values of the fusion rate constant $k_{u1}$ (solid lines), \change{which tunes between fragmented networks (blue), networks at the percolation transition (light green) and highly fused networks (yellow)}. 
	 The analytic approximation (Eq.~\ref{eq:slowdp}, dashed curves) encompases both the fractal continuum limit (for large clusters or fast decay) and the social network limit (for small clusters and slow decay). Plateaus at intermediate decay times correspond to filling of an individual cluster before interactions can occur.
	Insets show example simulation snapshots for the parameters indicated by the gray arrows. Parameters $N_0 = 250, \ell_0 = 0.5, R=5, k_{u2}=\frac{1}{3}k_{u1}, k_f=1$ are used in both (a) and (b), with the particle diffusivity set to $D_p=4800$ in (a) and $D_p=48$ in (b).	
	}
	\label{fig:alltogether}
\end{figure*}

\subsection*{Dynamic networks with large interacting clusters}
The model for spreading through a social network of interacting fragments can be expanded to approximate a regime with larger clusters. 
We use simulations of dynamic networks with different structures~\cite{holt2024spatiotemporal} to extract the effective parameters (mean cluster size $\left<n\right>$, effective unit diffusivity $D_n$, steric radius $a$, contact volume $v_c$, and effective local fusion rate $k_u$) for the simplified social network model (details in SI Appendix).
\change{The encounter time between clusters decreases with cluster size but remains well-approximated by the expression for diffusion-limited reactions of spherical particles, $\tau_\text{enc} = 1/(4\pi D b \rho)$~\cite{berg1977physics} (Supplemental Fig.~S2 in SI Appendix).}

When individual clusters are homogeneously filled upon each interaction ($\tau_c < \left\{\tau_d, \tau_\text{int}, \tau_f\right\}$), the solution in Eq.~\ref{eq:meanfieldsoln} can be used directly with the appropriate values of the parameters $a, v_c, k_u$. However, for slowly diffusing particles or very large clusters ($\tau_c > \tau_d$), the source cluster may be only partly full of material. 
The average concentration in the source cluster, $h_0^{(d)}$, is then determined by the static network solution (Eq.~\ref{eq:ddimsoln}) within the cluster: $\ell_0h_0^{(d)} =(S_d(\sqrt{D_p/k_d})+h_0\ell_0)/\langle n \rangle$. 
When another cluster fuses with the source cluster, 
fission may terminate the encounter before filling is complete
($\tau_f < \tau_c$). The amount transferred should then be scaled by the fraction of  the new cluster that is filled prior to fission: $f=(S_d(\sqrt{D_p/k_f})+h_0\ell_0)/\langle n \rangle$. These two corrections modify the dynamic equation for the material in the contact zone (Eq.~\ref{eq:meanfieldodes}b) and its solution (Eq.~\ref{eq:meanfieldsoln}a,b) as follows:
\begin{subequations}
	\begin{align}
		\frac{dh_c(t)}{dt} & = k_u \left[\ell_0 h_0^{(d)} -h_c(t) \right]f -\frac{I}{\rho v_c}-k_dh_c(t), \\
		S & = S_d\left(\sqrt{D_p/k_d}\right) + \rho h_c(v_c+z), \\
		h_c & =  \frac{\ell_0 h_0^{(d)}}{1+k_d/(f k_u)+k_dz/(fk_uv_c)},
	\end{align}
	\label{eq:slowdp}%
\end{subequations}
where the first term in the updated $S$ accounts for material in the source cluster and the second term represents material in the rest of the network. 
 When clusters are large and interactions are infrequent, the above solution reduces to the static network limit of Eq.~\ref{eq:ddimsoln}. When the clusters are small, the source cluster is fully filled ($h_0^{(d)}\rightarrow h_0$) and so is each cluster that interacts with it ($f \rightarrow 1$). The general solution then approaches the fragmented social network limit of Eq.~\ref{eq:meanfieldsoln}.

In Fig.~\ref{fig:alltogether}, we show the spreading of material through simulated networks with increasing connectivity, spanning across the different regimes. 
In the limit of rapid decay ($\tau_\textrm{d} < \tau_\textrm{c}$), material spreads within a single cluster of dimension $d$, with the \change{steady-state content} scaling as $S \sim \tau_d^{d/2}$.
When particle diffusivity is fast  and clusters are small, the source cluster fills faster than interactions can occur and the amount of material plateaus at the size of a single cluster, in the regime of $\tau_c < \tau_d < \tau_\text{int}$.
\change{This plateau implies that network filling is stalled over a certain timescale, consistent with past results on dynamic lattice networks below the percolation transition, when edge fluctuations are slow~\cite{hoitzing2015function,chuphal2024mitochondrial}. The plateau is expected to disappear when edges turn on and off rapidly, restoring network connectivity over longer timescales. In our system, this corresponds to longer decay times, for which interactions between distinct clusters dominate the spread. In this long-time regime, spreading through the population is three-dimensional and the total network content scales linearly as $S \sim \tau_d$. 

When particle diffusion is slow (Fig.~\ref{fig:alltogether}b), the cluster containing the source unit is able to offload material through interactions even before it is fully filled, and the plateau region narrows. If the rate of novel encounters is faster than the intra-cluster spreading timescale ($\tau_\text{int} < \tau_c$), the plateau disappears, as observed in lattice networks with rapid bond fluctuations~\cite{hoitzing2015function,chuphal2024mitochondrial}.}

The simplified analytic model (Eq.~\ref{eq:slowdp}) approximately matches simulations over a broad range of network connectivities and timescales (Fig.~\ref{fig:alltogether}). 
However, some discrepancy occurs for networks that are close to the percolation transition, particularly when the particle diffusivity is low. In this case, the approximation systematically overestimates material spreading through the network. This may be the result of transient fissions within a single cluster (which could temporarily hinder material spreading~\cite{chuphal2024mitochondrial}), or steric inaccessibility of nodes buried within a cluster, neither of which are accounted for in our analytic models. Additionally, the percolation transition corresponds to a broad variability in cluster sizes~\cite{sukhorukov2012emergence}, resulting in the source unit occasionally being trapped in very small clusters that limit material delivery.

\change{We note that in the limit of arbitrarily slow particle diffusivity or fragmented networks, mitochodrial content would remain limited to units which have come in direct contact with the source unit over a timescale of $\tau_d$. Steady-state filling is then determined by the cumulative degree of the temporal network describing all novel fusions made by a mitochondrial unit. This degree is plotted as a function of time in Supplemental Fig.~S2. Notably, the novel contact degree is highest for moderate cluster sizes, which enable rapid rearrangement of the network structure, as highlighted in prior work~\cite{holt2024spatiotemporal}. The ability of fragmented social networks of mobile mitochondria to efficiently share slowly-moving genomic elements has also been pointed out in the context of plant mitochondrial dynamics~\cite{giannakis2022exchange,chustecki2024collective}}.

The quantitative description of steady-state network filling (Eq.~\ref{eq:slowdp}) can be generalized to other measures of material spread on a dynamic network. In particular, the decay time $\tau_d$ sets the timescale over which the spreading is assessed. The total amount of material delivered into the network over time is also well-approximated by the physical and social network models presented here (see Supplemental Fig.~S3
in SI Appendix).

\subsection*{\label{sec:experiments} Spreading rates on mammalian mitochondrial networks}

\begin{figure*}
	\includegraphics[width=17.3cm]{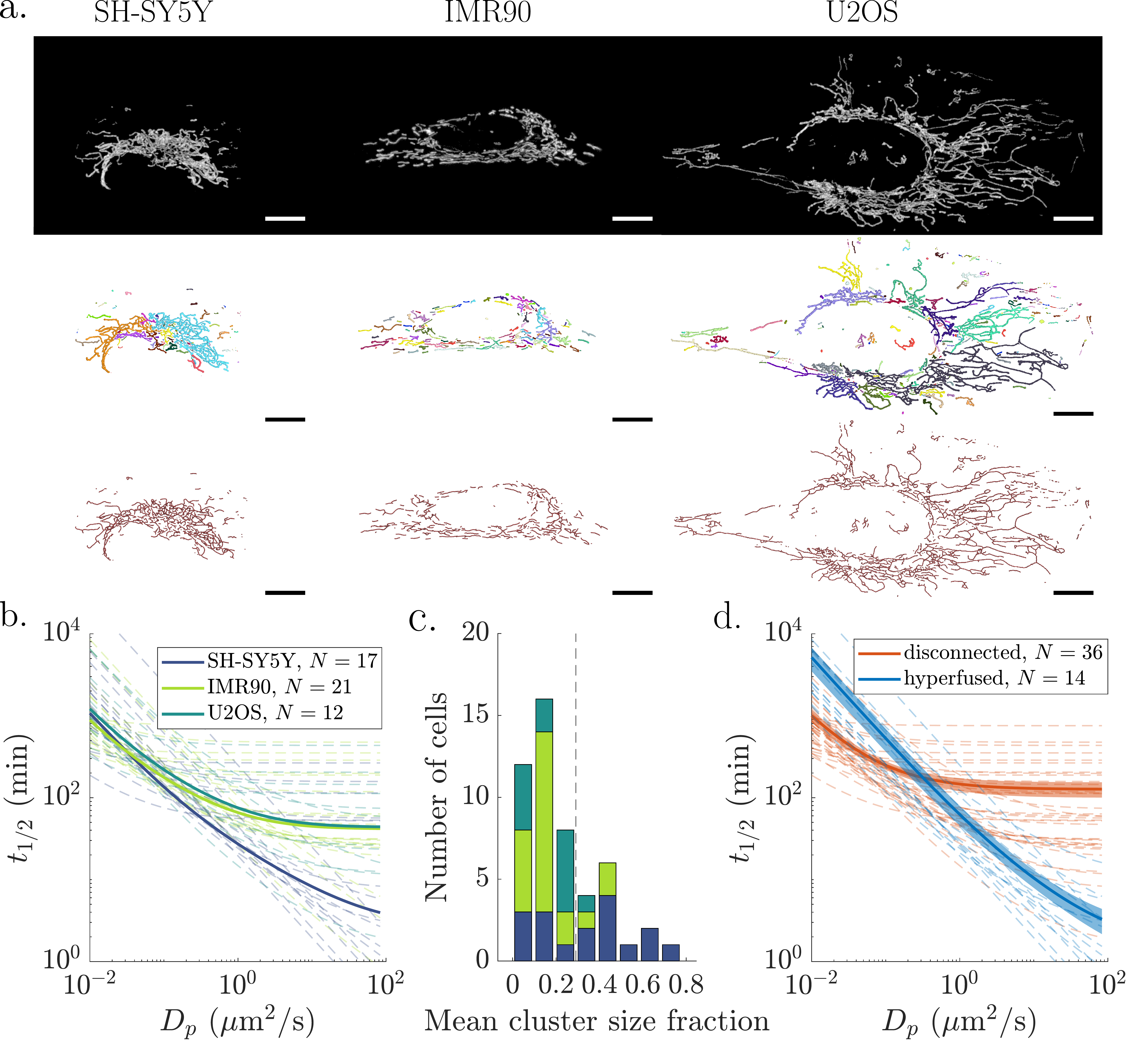}
	\caption{Extracted structures of mammalian mitochondrial networks exemplify both the static network and social network regimes. (a) Representative raw, segmented, and skeletonized images (top to bottom) for mitochondria from SH-SY5Y (human neuroblastoma), IMR90 (human fibroblast), and U2OS (human osteosarcoma) cells. Segmented images show each connected cluster with a different color. Scale bar: $10\mu\text{m}$. 
	(b) Time to fill half of the mitochondrial network with material supplied from a fixed source, as predicted by the analytic model, is plotted as a function of the material diffusivity for each cell type. Solid curves correspond to a `prototypical cell', where parameters are averaged within each cell type before plugging into the analytic model. Dashed lines indicate results for individual cells.
	(c) The distribution of cells by mean cluster size fraction (mean cluster length divided by total mitochondrial network length). The dashed vertical line indicates the threshold ($0.29$) to classify networks as disconnected or hyperfused, as determined by k-means clustering with k=2.	
	(d) Network filling half-times as in (b), with cells grouped by the mitochondrial network class (disconnected or hyperfused). Solid curves are obtained by averaging half-times across individual cells within each class with shaded regions indicating the standard error of the mean. Individual cell results are shown as dashed lines.
	}\label{fig:realcells}
\end{figure*}

To illustrate the utility of our model in providing directly testable predictions for mitochondrial mixing, we analyze time-resolved images of fluorescently tagged mitochondria in three human cell lines: SH-SY5Y, IMR90, and U2OS. Specifically, we segment, skeletonize, and track the mitochondria of each cell in our data set (Figure~\ref{fig:realcells}a and Supplemental Video 2), 
extracting structural metrics (degree distribution, cluster size, \change{graph} dimension) and dynamic parameters (fission rate and mitochondrial diffusivity). These values (provided in the SI Appendix) are used to parameterize our analytic model, which  provides a quantitative estimate of mitochondrial material spreading.
 We  estimate the time required for material to fill half the network ($t_{1/2}$) as the value of $\tau_d$ corresponding to $S$ reaching half its maximal value. This metric provides a prediction for how quickly particles of different types should spread through most of the mitochondrial network.

Figure~\ref{fig:realcells}b shows our predictions of spreading time on prototypical `average cells', with individual parameters averaged across all the cells of that type. The spreading time is plotted as a function of the material diffusivity, $D_p$, in real units. 
For typical mitochondrial matrix proteins with $D_p=20\mu \text{m}^2\text{/s}$~\cite{koopman2008inherited}, we expect a spreading time of $\sim 6$ min for SH-SY5Y, and $\sim 45$ min for IMR90 and U2OS cells.
Notably, the different cell types exhibit distinct behaviors. For IMR90 and U2OS cells, the spreading time asymptotes for particle diffusivities above roughly $1\mu\text{m}^2/\text{s}$. By contrast, in the typical SH-SY5Y cell, spreading times continue to decrease with higher material diffusivity. These results arise primarily from differences in the network architecture: SH-SY5Y mitochondria tend to be hyperfused, while IMR90 and U2OS exhibit more disconnected networks, so that dispersion becomes limited by interactions between distinct mitochondrial clusters rather than diffusion within connected mitochondria. 
Overall, our results indicate that cultured human cell lines can vary substantially in their mitochondrial architecture and \change{rate of material transport}.

It should be noted, however, that there is extensive variation in the mitochondrial structure and associated network parameters between individual cells of the same type (Fig.~\ref{fig:realcells}b). 
To further explore the dependence of spreading rates on network connectivity, we reclassified all cells in our data set as `disconnected' or `hyperfused' based on the mean cluster size as a fraction of the total mitochondrial population. Figure~\ref{fig:realcells}c shows the distribution of cells by mean cluster size fraction and the classification threshold  obtained via a k-means clustering algorthim. 
All cells with mean cluster size fraction below the threshold were classified as disconnected, with the remainder considered hyperfused. 
Figure~\ref{fig:realcells}d demonstrates the difference in spreading behavior for these two cell classes, with hyperfused cells exhibiting strong dependence on material diffusivity across the entire relevant range. Spreading on disconnected networks plateaus for particles with diffusivities above $D_p \approx 1\mu\text{m}^2\text{/s}$. We note that for slowly diffusing material, disconnected networks can actually increase the spreading rate \change{relative to hyperfused networks}. This is due 
to the mobility of mitochondrial clusters, which allows each unit to interact with more neighbors than they could contact in the  static low-dimensional hyperfused networks ($1 \leq d \lesssim 2$). \change{The enhanced rate of novel contacts between mitochondrial units in partially disconnected networks is demonstrated in the SI Appendix (Fig.~S2d)}.

The mitochondrial networks for different cell types exhibit multiple morphological differences. In particular, the total network size in U2OS cells is substantially larger than in the other two cell types (Fig.~\ref{fig:realcells}a; see SI Appendix). However, because the mean cluster size fraction is similar in both IMR90 and U2OS cells, the predicted filling half-time is similar as well. For systems that fall into the disconnected social network regime, it is the relative size of the individual mitochondrial clusters compared to the full network that determines filling times.

 The distinction in predicted spreading rates between different cell types can be attributed to their different propensities for mitochondrial connectivity. Many (though not all) SH-SY5Y cells exhibit hyperfused networks, while 
IMR90 and U2OS cells fall primarily in the disconnected category. \change{We note that these features may be dependent on the specific growth conditions (described in SI appendix) as well as the cell type.}
SH-SY5Y cells thus serve as examples of the physical network regime, where spreading is limited by material transport along a fractal-dimensional continuum. The other two cell types have mitochondria that behave akin to a social network, with interactions between clusters limiting the spread of rapidly diffusing particles.  Notably, for both IMR90 and U2OS mitochondrial networks, the spreading at high $D_p$ is limited by mitochondrial mobility rather than the timescale of local fusion or fission (see SI Appendix for extracted parameters). The balance of fusion to fission rates, which sets network connectivity, plays an important role in dictating material spread. However, our analysis indicates that the interaction rate is limited by mitochondria moving through space to encounter new fusion partners rather than by the waiting time for fusion between two nearby mitochondria ($ \tau_\text{enc} \gg \tau_u$). Thus, the model predicts that proportional increases in fusion and fission rates should have little effect on mitochondrial mixing.

\change{In other organisms and cell types, such as plant cells, mitochondria appear to be significantly more mobile, traversing tens of microns over a few-minute timescale~\cite{chustecki2021network}. Their active motion, which relies on processive motor-driven runs, results in rapid exploration throughout the cell, with encounter times apparently much smaller than the fusion timescale $ \tau_\text{enc} \ll \tau_u$ (see SI Appendix). In such a regime, mitochondrial material can be broadly spread out in space while individual units remain heterogeneous, as was observed for the mixing of photoconverted mitochondria in onion bulb epidermal cells~\cite{arimura2004frequent}. The biological variety of mitochondrial systems thus necessitates consideration of all the distinct timescales discussed in our general model for spreading through temporal networks.
}

\section*{\label{sec:discussion} Discussion}

In this work, we develop a mathematical framework for \change{spreading of diffusive material in temporal networks} and apply it to mitochondrial network structures in cultured human cells.
We examine two limiting regimes: one of static, well-connected physical networks and one of socially interacting  clusters.
In \change{static} networks, spreading rates increase with material diffusivity and network dimensionality. In three-dimensional social networks, material accumulates linearly in time and is limited either by the fusion rate between nearby clusters or the mobility of those clusters. \change{While the existence of these two extremes has been previously noted in models of epidemic spread among mobile agents~\cite{rodriguez2022epidemic}, we now provide analytic approximations that span over a broad range of intermediate connectivities, encompassing mass-conserving spreading of material through a network of dynamic clusters.}

The models presented here are based on measurable structural and dynamic features. The static model requires  network dimensionality, cluster size, and particle diffusivity, while the social model additionally needs the fusion rate and cluster mobility as inputs. Modern imaging techniques allow quantification of mitochondrial network structure and dynamics in a variety of cellular systems
~\cite{viana2015quantifying,wang2023mitotnt,zamponi2018mitochondrial,sukhorukov2012emergence,chen2009mitochondrial}. Our results connect these morphological measurements to the dynamics of material spreading through the mitochondrial population.

We show example calculations for three human cell types, demonstrating the distinct parameter regimes accessed by their mitochondrial networks. For the set of cells considered here, many SH-SY5Y cells fall in the well-connected regime, while the majority of IMR90 and U2OS cells exhibit disconnected social mitochondrial networks, whose mixing is limited by mitochondrial mobility and encounter. It should be noted, however, that the mitochondrial structure varies substantially among individual cells of a given type, and is also expected to change as a function of metabolic conditions~\cite{liesa2013mitochondrial} and genetic perturbations~\cite{zamponi2018mitochondrial,izzo2017metformin}.
The predictions made by this model could be tested in future work by quantifying the dispersion of locally photoconverted proteins through mitochondrial networks of different architectures~\cite{liesa2013mitochondrial}.

Our quantitative calculations focus on a simplified dynamic system, with a single source of material held at fixed concentration. Realistic biological scenarios are, of course, likely to involve additional complications, including multiple sources (such as the mitochondrial nucleoids), non-trivial regulation at the source, or diffusive barriers within the mitochondria~\cite{dieteren2011solute,jakubke2021cristae}. We also assumed that individual mitochondrial units are all identical in their fusion and fission behavior, whereas it is possible that some units remain isolated from the rest of the network, while others are more likely to engage in interactions. Dynamic rearrangements beyond fusion and fission (such as tubule extension and branch sliding)  may also contribute to \change{material transport} within individual mitochondrial network components~\cite{lewis2023mitochondrial,osman2015integrity,wang2015dynamic}. \change{Furthermore, mitochondria have been observed to engage in directed motor-driven motion in both plant~\cite{chustecki2021network} and mammalian~\cite{winter2025asymmetric} cells, likely enhancing the rate of encounter in the fragmented regime (see SI Appendix).}
However, the minimal model presented here provides a basic building block from which more complicated scenarios can be constructed and which can serve as a null hypothesis for analyzing experimental observations. 

\change{This study focuses on dispersion through the mitochondrial population. A potentially fruitful avenue of future work would be exploring how mitochondrial material is dynamically distributed throughout the cellular space. Developing a mathematically tractable approach to this question would require establishing the extrinsic spatial dimensionality~\cite{ben2000diffusion,meakin1984topological} of mitochondrial clusters, as well as accounting for heterogeneous transport of the organelles~\cite{schwarz2016optimality,giannakis2022exchange} and for the varying geometry of the intracellular space. 
	Such studies could further illuminate how different mitochondrial architectures resolve the tension between cellular functions that require mitochondria to be broadly dispersed yet well-mixed in their complement of proteins, RNAs, and genetic components~\cite{chustecki2024collective}.}

This work brings us closer to understanding the functional implications of variable network connectivity for the spread of ions, lipids, proteins, and genetic information within mitochondria. 
 For example, diffusive spreading dictates the heterogeneity of mitochondrial contents arising from mRNA and protein import site localization on the mitochondrial surface~\cite{arceo2022mitochondrial,khan2024mitochondrial}. Similarly,  diffusion of mitochondrial transcription factors and mRNAs may dictate the spheres of influence of individual nucleoids within the network~\cite{jakubke2021cristae}.
 
Mitochondrial networks are known to alter their architecture in response to changing environmental conditions, with excess glucose causing network fragmentation and starvation resulting in hyperfused networks~\cite{liesa2013mitochondrial}. A quantitative framework for material spreading in different structures provides insight on the implications of these transitions for many biologically critical processes, such as the distribution of harmful reactive oxygen species~\cite{yu2006increased}, the ability to isolate or complement deleterious DNA mutations~\cite{nakada2001inter,jakubke2021cristae}, and the tunneling of calcium ions, proton gradients, and ATP throughout the cell~\cite{bravo2017calcium,hoitzing2015function,skulachev2001mitochondrial}

While we have focused on mitochondria, analogous problems of diffusive spreading arise in many systems, such as porous media in the geosciences~\cite{herrera2013fractal}, trophallactic networks in honeybees~\cite{gernat2018automated}, and Alzheimers progression across connected regions in the brain~\cite{raj2015network}. By bridging across models of static physical networks and transiently interacting social networks, our results enable a quantitatively predictive link from \change{temporal} network \change{structure} to the rate of diffusive dispersion.

\section*{Materials and Methods}

\subsection*{Mathematical derivations}
The exact solution and the fractal-continuum approximation for filling of highly connected stationary networks (Eq.~\ref{eq:ddimsoln}, Fig.~\ref{fig:ddimmatchescontinuum}) were derived as shown in sections~1A,1B
 of the SI Appendix. The analytic approximation for the social network model (Eq.~\ref{eq:meanfieldodes},\ref{eq:meanfieldsoln}, Fig.~\ref{fig:meanfieldsoln}) was derived as shown in  section~1C
 of the SI Appendix.

\subsection*{Simulation methods}
Agent-based spatially resolved simulations of dynamic networks consisting of spherocylindrical units undergoing fusion and fission were carried out as described in prior work~\cite{holt2024spatiotemporal}. Spreading of material within these dynamic networks was modeled using a finite volume method, with simulation details provided in section~1D
of the SI Appendix. \change{Simulation code and example parameter files are available at \url{https://github.com/lenafabr/mitochondrialNetworks}}.

\subsection*{Cell culture and imaging methods}
Cell culture, microscopy, mitochondrial network extraction, and measurements of dynamic parameters were carried out as described in section~1E
of the SI Appendix. Specific plasmids used are provided in table~S1
and extracted averaged parameters are given in table~S2.
The mitochondrial network model was modified to use a slab-like  geometry for analysis of spreading in experimentally extracted network structures, as described in section~1E
of the SI Appendix. \change{Imaging data is available at \url{https://doi.org/10.5061/dryad.cjsxksnkb}}.

\section*{Acknowledgements}
{EFK acknowledges funding from NSF grant PHYS-2310229, the Chan Zuckerberg Initiative, and a grant from the UCSD Academic Senate. SL acknowledges funding by CZI Metabolism Across Scales grant \#RR-8817, and by NSF CAREER award \#2339182. We are grateful to Austin EYT Lefebvre, Gabriel Sturm, Nigel Goldenfeld, and Suliana Manley and to the members of the CZI Theory Institute Without Walls for helpful discussions.}

\bibliography{spreadondynamicnetworks}

\end{document}